\documentclass[a4paper,11pt]{llncs}
\usepackage[english]{babel}
\usepackage{german}
\usepackage{amssymb}
\usepackage{amsfonts}
\usepackage{amsmath}
\usepackage[latin1]{inputenc}
\usepackage{fullpage}
\usepackage{cite}
\usepackage{graphicx}
\usepackage{subfigure}
\usepackage{algorithm}
\usepackage[noend]{algorithmic}
\usepackage{amsmath,amssymb,cite}
\usepackage{footmisc,marvosym}
\usepackage{wasysym,vmargin,stackrel}

\begin{document}

\mainmatter              
\title{Maximum Cliques in Graphs with Small Intersection Number and Random Intersection Graphs}
%
%
\author{S. Nikoletseas\inst{1,2} 
\and C. Raptopoulos\inst{1}
\and P. G. Spirakis \inst{1,2}}
%
%
%
\institute{Computer Technology Institute, P.O. Box 1122, 26110 Patras, Greece\\
\and University of Patras, 26500 Patras, Greece \\
\email{nikole@cti.gr}, \email{raptopox@ceid.upatras.gr}, \email{spirakis@cti.gr}}

\maketitle              

\begin{abstract}
In this paper, we relate the problem of finding a maximum clique to the intersection number of the input graph (i.e. the minimum number of cliques needed to edge cover the graph). In particular, we consider the maximum clique problem for graphs with small intersection number and random intersection graphs (a model in which each one of $m$ labels is chosen independently with probability $p$ by each one of $n$ vertices, and there are edges between any vertices with overlaps in the labels chosen). 

We first present a simple algorithm which, on input $G$ finds a maximum clique in $O(2^{2^m + O(m)} + n^2 \min\{2^m, n\})$ time steps, where $m$ is an upper bound on the intersection number and $n$ is the number of vertices. Consequently, when $m \leq \ln{\ln{n}}$ the running time of this algorithm is polynomial.

We then consider random instances of the random intersection graphs model as input graphs. As our main contribution, we prove that, when the number of labels is not too large ($m=n^{\alpha}, 0< \alpha <1$), we can use the label choices of the vertices to find a maximum clique in polynomial time whp. The proof of correctness for this algorithm relies on our Single Label Clique Theorem, which roughly states that whp a ``large enough'' clique cannot be formed by more than one label. This theorem generalizes and strengthens other related results in the state of the art, but also broadens the range of values considered (see e.g. \cite{S95} and \cite{BTU08}).

As an important consequence of our Single Label Clique Theorem, we prove that the problem of inferring the complete information of label choices for each vertex from the resulting random intersection graph (i.e. the \emph{label representation of the graph}) is \emph{solvable} whp; namely, the maximum likelihood estimation method will provide a unique solution (up to permutations of the labels). Finding efficient algorithms for constructing such a label representation is left as an interesting open problem for future research.
 
\end{abstract}

\section{Introduction}

A \emph{clique} in an undirected graph $G$ is a subset of vertices any two of which are connected by an edge. The cardinality of the maximum clique is called the \emph{clique number} of $G$. The problem of finding the maximum clique in an arbitrary graph is fundamental in Theoretical Computer Science and appears in many different settings. As an example, consider a social network where vertices represent people and edges represent mutual acquaintance. Finding a maximum clique in this network corresponds to finding the largest subset of people who all know each other. More generally, the analysis of large networks in order to identify communities, clusters,
and other latent structure has come to the forefront of much research. The Internet, social networks, bibliographic databases, energy distribution networks, and global networks of economies are some of the examples motivating the development of the field.

It is well known that determining the clique number of an arbitrary graph is NP-complete \cite{K72}. In fact, the fastest algorithm known today runs in time $O(1.1888^n)$ \cite{R01}, where $n$ is the number of vertices in the graph. Moreover, the best known approximation algorithm for the clique number has a performance guarantee of $O\left( \frac{n (\log\log{n})^2}{(\log{n})^3}\right)$ \cite{F04} (there are algorithms with better approximation ratios for graphs with large clique number; see e.g. \cite{AK98}). Even though this approximation ratio appears to be weak at first glance, there are several results on hardness of approximation which suggest that there can be no approximation algorithm with an approximation ratio significantly less than linear (see e.g. \cite{H99}). It was also shown in \cite{CHKX06} that, if $k$ is the clique number, then the clique problem cannot be solved in time $n^{o(k)}$, unless the exponential time hypothesis fails (note that the brute force search algorithm runs in time $O(n^k k^2)$, which seems quite close).

The intractability of the maximum clique problem for arbitrary graphs lead researchers to the study of the problem for appropriately generated random graphs. In particular, for Erd\H{o}s-R\'enyi random graphs $G_{n, \frac{1}{2}}$ (i.e. random graphs in which each edge appears independently with probability $\frac{1}{2}$), there are several greedy algorithms that find a clique of size about $\ln{n}$ with high probability (whp, i.e. with probability that tends to 1 as $n$ goes to infinity), see e.g. \cite{GM75, K76}. Since the clique number of $G_{n, \frac{1}{2}}$ is asymptotically equal to $2 \ln{n}$ with high probability, these algorithms approximate the clique number by a factor of 2. In fact, it was conjectured that finding a clique of size $(1+\epsilon) \ln{n}$ (for a constant $\epsilon>0$), with probability at least $\frac{1}{2}$, would require techniques beyond the current limits of complexity theory. This belief was strengthened by the fact that the Metropolis algorithm also fails to find the maximum clique in $G_{n, \frac{1}{2}}$ (see \cite{J92}). A more dramatized version of the above conjecture was presented in \cite{J92}, stating that the problem of finding an $1.01 \ln{n}$ clique remains hard even if the input graph is a $G_{n, \frac{1}{2}}$ random graph in which we have planted a randomly chosen clique of size $n^{0.49}$. This conjecture has some interesting cryptographic consequences, as shown in \cite{JP98}. It also seems tight, since finding the maximum clique in the case where the planted clique has size at least $\sqrt{n}$ can be done in polynomial time by using spectral properties of the adjacency matrix of the graph (see \cite{AKS98}). We finally note that there are quite a few nice results concerning generalizations of the planted clique problem in various (quite general) random graphs models (see e.g. \cite{C06, CL09}).

\subsection{Our Contribution}

In this work, we complement the state of the art by relating the maximum clique problem to the intersection number of the input graph $G$ (i.e. the minimum number of cliques that can edge cover $G$). In particular, we consider the maximum clique problem for graphs with small intersection number and random intersection graphs.

More analytically, we begin by considering arbitrary graphs with small intersection number. We present a simple algorithm which, on input $G$ finds a maximum clique in $O(2^{2^m + O(m)} + n^2 \min\{2^m, n\})$ time steps, where $m$ is an upper bound on the intersection number of $G$ and $n$ is the number of vertices. Consequently, when $m \leq \ln{\ln{n}}$ the running time of this algorithm is polynomial. We note here that computing the exact value of the independence number of $G$ is itself an NP-complete problem, but this knowledge is only needed in the analysis of the algorithm. 

We then consider random instances of the random intersection graphs model (introduced in \cite{KSS99, S95}) as input graphs. In this model, denoted by ${\cal G}_{n, m, p}$, each one of $m$ labels is chosen independently with probability $p$ by each one of $n$ vertices, and there are edges between any vertices with overlaps in the labels chosen. Random intersection graphs are relevant to and capture quite nicely social networking. Indeed, a social network is a structure made of nodes (individuals or organizations) tied by one or more specific types of interdependency, such as values, visions, financial exchange, friends, conflicts, web links etc. Social network analysis views social relationships in terms of nodes and ties. Nodes are the individual actors within the networks and ties are the relationships between the actors. Other applications include oblivious resource sharing in a (general) distributed setting, efficient and secure communication in sensor networks \cite{NRS11}, interactions of mobile agents traversing the web etc. Even epidemiological phenomena (like spread of disease) tend to be more accurately captured by this ``interaction-sensitive'' random graph model.

As our main contribution, we prove that, when the number of labels is not too large, we can use the label choices of the vertices to find a maximum clique in polynomial time (in the number of labels $m$ and vertices $n$ of the graph). Most of the work in this paper is devoted in proving our Single Label Clique Theorem (Theorem \ref{singlelabelclique} in Section \ref{sectionclique}). Our proof technique is original and employs a probabilistic contradiction argument. The theorem states that when the number of labels is less than the number of vertices, any large enough clique in a random instance of ${\cal G}_{n, m, p}$ is formed by a single label. This statement may seem obvious when $p$ is small, but it is hard to imagine that it still holds for \emph{all} ``interesting'' values for $p$ (see also the discussion in Section \ref{definitionsnotationrange}). Indeed, when $p = o\left(\sqrt{\frac{1}{nm}} \right)$, by slightly modifying an argument of \cite{BTU08}, we can see that $G_{n, m, p}$ almost surely has no cycle of size $k \geq 3$ whose edges are formed by $k$ distinct labels (alternatively, the intersection graph produced by reversing the roles of labels and vertices is a tree). On the other hand, for larger $p$ a random instance of ${\cal G}_{n, m, p}$ is far from perfect\footnote{A \emph{perfect graph} is a graph in which the chromatic number of every induced subgraph equals the size of the largest clique of that subgraph. Consequently, the clique number of a perfect graph is equal to its chromatic number.} and the techniques of \cite{BTU08} do not apply (for a more thorough discussion see the beginning of Section \ref{sectionclique}). By using the Single Label Clique Theorem, we provide a tight bound on the clique number of $G_{n, m, p}$ when $m = n^{\alpha}, \alpha<1$. A lower bound in the special case where $mp^2$ is constant, was given in \cite{S95}. We considerably broaden this range of values to also include vanishing values for $mp^2$ and also provide an asymptotically tight upper bound. 

We claim that our proof also applies for $\alpha<2$, provided $p$ is not too small. We should note here that in \cite{FSS00} the authors prove the equivalence (measured in terms of total variation distance) of random intersection graphs and Erd\H{o}s-R\'enyi random graphs, when $m = n^{\alpha}, \alpha>6$. This bound on the number of labels was improved in \cite{R11}, by showing equivalence of sharp threshold functions among the two models for $\alpha \geq 3$. In view of these results, we expect that our work will shed light also in the problem of finding maximum cliques in Erd\H{o}s-R\'enyi random graphs.

Finally, as yet another consequence of our Single Label Clique Theorem, we prove that the problem of inferring the complete information of label choices for each vertex from the resulting random intersection graph (i.e. the \emph{label representation of the graph}) is \emph{solvable} whp; namely, the maximum likelihood estimation method will provide a unique solution (up to permutations of the labels).\footnote{More precisely, if ${\cal B}$ is the set of different label choices that can give rise to a graph $G$, then the problem of inferring the complete information of label choices from $G$ is \emph{solvable} if there is some $B^* \in {\cal B}$ such that $\Pr(B^*|G) > \Pr(B|G)$, for all ${\cal B} \ni B \neq B^*$.} In particular, given values $m, n$ and $p$, such that $m = n^{\alpha}, 0<\alpha<1$, and given a random instance of the ${\cal G}_{n, m, p}$ model, the label choices for each vertex are uniquely defined. Finding efficient algorithms for constructing such a label representation is left as an open problem for future research.

\subsection{Organization of the paper}

In Section \ref{definitionsnotationrange} we formally define random intersection graphs. We also provide some useful definitions and notation which are used throughout the paper. The relation of the intersection number to the clique number of an arbitrary graph is discussed in Section \ref{aritrarymaxclique}. Section \ref{sectionclique} is devoted to the proof of our Single Label Clique Theorem for random intersection graphs. The consequences of our main theorem concerning the efficient construction of a maximum clique and the uniqueness of the label representation of $G_{n, m, p}$ are presented in Section \ref{labelreconstruction}. Finally, we discuss the presented results and further research in Section \ref{conclusions}.

\section{Definitions and Preliminaries} \label{definitionsnotationrange}

The formal definition of the random intersection graphs model is as follows:

\begin{definition}[Random Intersection Graph - ${\cal G}_{n, m, p}$ \cite{KSS99,S95}]
Consider a universe ${\cal M} = \{1, 2, \ldots, m\}$ of elements and a set of $n$ vertices $V$. Assign independently to each vertex $v \in V$ a subset $S_{v}$ of ${\cal M}$, choosing each element $i \in {\cal M}$ independently with probability $p$ and draw an edge between two vertices $v \neq u$ if and only if $S_{v} \cap S_{u} \neq \emptyset$. The resulting graph is an instance $G_{n, m, p}$ of the random intersection graphs model. 
\end{definition}

In this model we also denote by $L_i$ the set of vertices that have chosen label $i \in M$. Given $G_{n, m, p}$, we will refer to $\{L_{i}, i \in {\cal M}\}$ as its \emph{label representation}. Consider the bipartite graph with vertex set $V \cup {\cal M}$ and edge set $\{(v, i): i \in S_{v}\} = \{(v, i): v \in L_i\}$. We will refer to this graph as the \emph{bipartite random graph $B_{n, m, p}$ associated to $G_{n, m, p}$}. Notice that the associated bipartite graph is uniquely defined by the label representation.

It follows from the definition of the model that the edges in $G_{n, m, p}$ are not independent. In particular, the (unconditioned) probability that a specific edge exists is $1-(1-p^2)^m$. Therefore, if $mp^2$ goes to infinity with $n$, then this probability goes to 1. In the paper, we will thus consider the ``interesting'' range of values $mp^2 = O(1)$ (i.e. the range of values for which the unconditioned probability that an edge exists does not go to 1). Furthermore, as is usual in the literature, we will assume that the number of labels is some power of the number of vertices, i.e. $m = n^{\alpha}$, for some $\alpha >0$.

The following definitions will also be useful:

\begin{definition}[Intersection number]
The intersection number of a graph $G$ is the smallest number of cliques needed to cover all of the edges of $G$.
\end{definition}
Equivalently, the intersection number is the smallest number of elements in a representation of $G$ as an intersection graph of finite sets.

\begin{definition}[Edge clique cover]
A set of cliques ${\cal C} = \{C_1, \ldots, C_m\}$ is an edge clique cover of a graph $G = (V, E)$ if for every edge $e \in E$ there is at least one clique $C_i$ such that $e \in C_i$ and for every non edge $e' \notin E$, there is no such clique in ${\cal C}$. 
\end{definition}
Therefore, the intersection number of $G$ is the minimum $m$ such that ${\cal C} = \{C_1, \ldots, C_m\}$ is an edge clique cover of $G$.

\subsection{Notation}

We use the convention that the random intersection graphs model is denoted by ${\cal G}_{n, m, p}$ (i.e. with a calligraph ${\cal G}$), while a specific random instance of the model is denoted by $G_{n, m, p}$ (i.e. with a simple $G$). 

For a vertex $v \in V$, we denote by $N_G(v)$ the set of neighbors of $v$ in $G$. We will say that two vertices $v, u \in V$ belong to the same \emph{closed neighborhood in $G$} and we will write $v \leftrightarrow_G u$ if and only if $N_G(v) \cup \{v\} = N_G(u) \cup \{u\}$. 

Let ${\cal C}'$ denote a partition of the vertex set $V$ of a graph $G$ and let $v \in V$. We will denote by ${\cal C}'[v]$ the unique set inside ${\cal C}'$ that contains $v$, that is ${\cal C}'[v] = \{C' \in {\cal C}': v \in C'\}$. 

Throughout the paper, we make use of the well known asymptotic notation $O(\cdot), \Omega(\cdot), o(\cdot)$ and $\omega(\cdot)$. Furthermore, we use the relation ``$\sim$'' for asymptotically equal. In particular, if $f(n), g(n)$ are two functions of $n$, then $f(n) \sim g(n)$ means that $\lim_{n \to \infty} \frac{f(n)}{g(n)} = 1$ or equivalently $f(n) = g(n) + o(g(n))$.

\section{An Algorithm for Maximum Clique} \label{aritrarymaxclique}

In this section we consider arbitrary graphs as input graphs for the maximum clique problem. In particular, we relate the running time of the following algorithm to the intersection number of the input graph $G$.

\vspace{0.5cm}

\begin{tabular}{|p{12cm}|}
\hline
\noindent \texttt{\textbf{Algorithm}} FIND\_MAX-CLIQUE \\
\noindent \texttt{\textbf{Input:}} $G = (V, E)$

\begin{enumerate}

\item Set $U = V$ and ${\cal C}' = \emptyset$; 

\% Form the closed neighborhood partition \%

\item \texttt{\textbf{while}} $U \neq \emptyset$ \texttt{\textbf{do}} 

\item \quad Pick $v \in U$ and let $C' = \{u \in U: u \leftrightarrow_G v\}$; 

\item \quad Include $C'$ in ${\cal C}'$;  

\item \quad Set $U = U \backslash C'$; \texttt{\textbf{endwhile}}

\% Define an induced subgraph \%

\item Let $G' = (V', E')$ be an induced subgraph of $G$ that has exactly one vertex for every set $C' \in {\cal C}'$;

\% Find a clique of $G'$ that corresponds to the maximum clique of $G$ \%

\item Using exhaustive search, find a clique $S$ in $G'$ such that $|\cup_{v' \in S} {\cal C}'[v']|$ is maximum; 

\item \texttt{\textbf{Output}} $Q = \cup_{v' \in S} {\cal C}'[v']$;

\end{enumerate} \\
\hline 
\end{tabular}

\vspace{0.5cm}

\begin{figure}[htb]
\centering 
\includegraphics[scale=0.7]{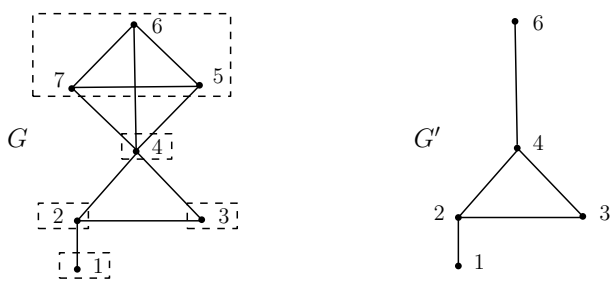} 
\caption{An example of a graph $G$ and corresponding $G'$.}
\label{Gintersectionmaxclique}
\end{figure}

An example of how the graph $G'$ is constructed (in step 6) for a specific graph $G$ is shown in Figure \ref{Gintersectionmaxclique}. Notice that $G$ has five closed neighborhoods (whereas its intersection number is 3), which are shown in dashed squares, so the graph $G'$ has 5 vertices. The corresponding clique of $G'$ that maximizes $|\cup_{v' \in S} \{u:u \leftrightarrow_G v'\}|$ is $S = \{4, 6\}$.

\subsection{Analysis of  FIND\_MAX-CLIQUE}

We first present the following lemma that concerns basic properties of the relation $\leftrightarrow_G$.

\begin{lemma} \label{relationproperties}
The closed neighborhood relation $\leftrightarrow_G$ is an equivalence relation with the following properties: 

\begin{enumerate}

\item It is an equivalence relation which partitions the vertex set $V$ in equivalence classes called closed neighborhoods. 

\item A closed neighborhood is a clique. Two closed neighborhoods either form a clique, or no edge between their vertices exists. 

\end{enumerate}  
\end{lemma} 
\emph{Proof.} (1) The fact that $\leftrightarrow_G$ is an equivalence relation follows directly by its definition. Therefore, every vertex belongs to exactly one equivalence class (i.e. exactly one closed neighborhood).

(2) By definition, a closed neighborhood forms a clique. Let now $C_1', C_2'$ be two distinct closed neighborhoods and let $v \in C_1', u \in C_2'$. Suppose that there is an edge $(u, v) $ between $u$ and $v$ in $G$, i.e. $u \in N_G(v)$. Consider now \emph{any} two vertices $v' \in C_1', u' \in C_2'$ (including $v, u$). By definition of the closed neighborhood relation, we must have that $N_G(v) \cup \{v\} = N_G(v') \cup \{v'\}$. Since the close neighborhoods are disjoint, this means that $u' \in N_G(v')$. Therefore, either every edge between $C_1'$ and $C_2'$ appears in $G$, and $C_1' \cup C_2'$ forms a clique, or no edge between them exists. This completes the proof.

$\hfill \Box$

We now prove the following theorem about the correctness of the Algorithm  FIND\_MAX-CLIQUE.

\begin{theorem}[Correctness]
FIND\_MAX-CLIQUE correctly outputs a maximum clique in $G$.
\end{theorem}
\emph{Proof.} Notice that, by the second part of Lemma \ref{relationproperties} and by construction of $G'$, any clique $S$ in $G'$ corresponds to a clique $\cup_{v' \in S} {\cal C}'[v']$ in $G$. 

Therefore, we only need to show that a maximum clique $Q$ of $G$ corresponds to a clique in $G'$, because then the algorithm will be able to find it in step 7. Equivalently, we need to show that there are $k \geq 1$ closed neighborhoods $C_1', \ldots, C_k'$ which constitute a partition of $Q$, that is $\cup_{i=1}^k C_i' = Q$. Indeed, by construction of $G'$, the vertices in $G'$ that correspond to these closed neighborhoods will form a clique in $G'$ (any choice of two vertices will be connected). 

To prove the above, let $C'$ be a closed neighborhood that has at least one common vertex $v'$ with $Q$, i.e. $v' \in C' \cap Q$. Then, by definition of the $\leftrightarrow_G$ relation, every vertex $u' \leftrightarrow_G v'$ is connected to $v'$ and to all the vertices that $v'$ is connected to (including all vertices in $Q$). Therefore, by maximality of $Q$, all the vertices in $C'$ must be contained in the maximum clique, i.e. $C' \subseteq Q$. Consequently, a closed neighborhood is either entirely contained in $Q$, or disjoint from it. By the first part of Lemma \ref{relationproperties}, we can then partition $Q$ using all the closed neighborhoods that have common vertices with $Q$. This completes the proof. 

$\hfill \Box$

The following result relates the running time of Algorithm  FIND\_MAX-CLIQUE to the intersection number of its input graph $G$.

\begin{theorem}[Efficiency] \label{efficiency}
Let $G = (V, E)$ be a graph with intersection number $m$. Then FIND\_MAX-CLIQUE on input $G$ finds a maximum clique in $O(2^{2^m + O(m)} + n^2 \min\{2^m, n\})$ time steps.
\end{theorem}
\emph{Proof.} By definition, since the intersection number of $G$ is $m$, there is a set of cliques ${\cal C} = \{C_1, \ldots, C_m\}$ that is an edge clique cover of $G$. For a vertex $v \in V$, we denote by $S_v$ the set of cliques in ${\cal C}$ that include $v$. Notice then that if $S_v = S_u$, then not only are $u$ and $v$ connected, but they also have the exact same set of neighbors in $V \backslash \{u, v\}$, i.e. $N_G(v) \cup \{v\} = N_G(u) \cup \{u\}$. 

Given now a specific edge clique cover ${\cal C}$, there are at most $2^m$ different ways in which we can construct a set $S_v$. Consequently, there are at most $2^m \leq n$ distinct closed neighborhoods $C_1', \ldots, C_{2^m}'$ in $G$ which constitute a partition of the set of non-isolated vertices. Note also that determining whether or not $v \leftrightarrow_G u$ for any two vertices requires $O(n)$ steps. Therefore, steps 2 to 5 needed for partitioning the vertex set $V$ in closed neighborhoods in the algorithm require $O(n^2 \min\{2^m, n\})$ time.

From the above, we also conclude that the number of vertices in $G'$ is at most $2^m$. Therefore, the time needed to construct $G'$ in step 6 in the algorithm is $O(2^{2m})$. Finally, there are at most $2^{2^m}$ subsets of vertices in $G'$, so step 7 in the algorithm takes $O(2^{2^m+2m})$ time. This completes the proof.

$\hfill \Box$

Note that the algorithm does not need the actual value of the independence number. We only use this information for bounding its running time. The following is a direct consequence of Theorem \ref{efficiency}.

\begin{corollary}
Let $m \leq \ln{\ln{n}}$ be an upper bound on the independence number of an arbitrary undirected graph $G$ on $n$ vertices. Then there is an algorithm that finds the maximum clique of $G$ in time $O(n^2 \ln{n})$. 
\end{corollary}

As a final remark, since the intersection number of $G_{n, m, p}$ is at most $m$ (but could be even less), the above result also holds for any random instance of the random intersection graphs model with at most $\ln{\ln{n}}$ labels.


\section{Clique number for $m = n^{\alpha}, 0<\alpha<1$} \label{sectionclique}

In this section we give a tight bound on the clique number of $G_{n, m, p}$ when $m = n^{\alpha}, \alpha<1$. A lower bound in the special case where $mp^2$ is constant, was given in \cite{S95}. We considerably broaden this range of values to also include vanishing values for $mp^2$ and also provide a tight upper bound. 

We will also assume, without loss of generality, that $p = \Omega\left(\sqrt{\frac{1}{nm}} \right)$. Indeed, when $p = o\left(\sqrt{\frac{1}{nm}} \right)$, by slightly modifying an argument of \cite{BTU08}, we can see that $G_{n, m, p}$ almost surely has no cycle of size $k \geq 3$ whose edges are formed by $k$ distinct labels. Therefore, the maximum clique of $G_{n, m, p}$ when $p = o\left(\sqrt{\frac{1}{nm}} \right)$, is formed by exactly one label. As a matter of fact, if $L_i$ is the set of vertices that have chosen label $i \in {\cal M}$, then the maximum clique is equal to $L_l$, where $l \in \arg\max_{i \in {\cal M}}|L_i|$. Furthermore, since $G_{n, m, p}$ is chordal whp (see Lemma 5 in \cite{BTU08}), the maximum clique can be found in polynomial time. 

We stress out the fact that the techniques employed to provide the algorithmic and structural results in \cite{BTU08} cannot be used in the case where $p = \Omega\left(\sqrt{\frac{1}{nm}} \right)$. In particular, $G_{n, m, p}$ is far from perfect, especially in the the case $mp = \omega(\ln{n})$ (which is included in the range of values that we study here). An intuitive justification is as follows: when $mp = \omega(\ln{n})$, then the size of the label sets of every vertex are highly concentrated around their mean value $mp$. Therefore, the statistical behavior of $G_{n, m, p}$ is expected to be similar to the statistical behavior of uniform random intersection graphs $G_{n, m, \lambda}$, in which each vertex selects exactly $\lambda = mp$ labels from ${\cal M}$. It was proved in \cite{TCS10} (part (iii) in Corollary 2), that the size of the maximum independent set when $m = n^{\alpha}, \alpha<1$ and $\lambda = \omega(\ln{n})$, is asymptotically equal to $2 (1-\alpha) \frac{m \ln{n}}{\lambda^2}$. Therefore, when $mp = \omega(\ln{n})$, the size of the maximum independent set in $G_{n, m, p}$ will be around $\Theta\left(\frac{\ln{n}}{mp^2} \right)$, so its chromatic number will be $\Omega\left(\frac{nmp^2}{\ln{n}}\right)$. However, as can be seen in Corollary \ref{maxcliquesize} (which is a direct consequence of our main theorem), the size of the maximum clique in $G_{n, m, p}$ when $m = n^{\alpha}, \alpha<1$ and $mp^2 = O(1)$ is asymptotically equal to $np$. This is much smaller than the lower bound $\Omega\left(\frac{nmp^2}{\ln{n}}\right)$ on the chromatic number in the case $mp = \omega(\ln{n})$. Therefore, $G_{n, m, p}$ is far from perfect in this range of values.

We first provide some concentration results concerning the number of vertices that have chosen a particular label and the number of vertices that have chosen two particular labels.

\begin{lemma} \label{lemmaconcentration}
Let $G_{n, m, p}$ be a random instance of the random intersection graphs model with $m = n^{\alpha}, 0<\alpha<1$ and $p = \Omega\left(\sqrt{\frac{1}{nm}} \right)$. Then the following hold:

\begin{description}

\item[A.] Let $L_i$ be the set of vertices that have chosen label $i \in {\cal M}$. Then 

\begin{equation} 
\Pr(\exists i \in {\cal M}:||L_i| - np| \geq 3 \sqrt{np \ln{n}}) \leq \frac{1}{n^3} \to 0.
\end{equation}

\item[B.] Let also $S_v$ denote the set of labels that were chosen by vertex $v$. Then 

\begin{equation} 
\Pr(\exists v \in V:|S_v| > mp + 3 \sqrt{mp \ln{m}} + \ln{n}) \to 0.
\end{equation}

\end{description}
\end{lemma}
\proof For the first part, fix a label $i \in {\cal M}$. Notice that $|L_i|$ is a binomial random variable with parameters $n, p$, i.e. $|L_i| \sim {\cal B}(n, p)$. By Chernoff bounds, for any $t \geq 0$, we have that

\begin{displaymath}
\Pr(||L_i| - np| \geq t) \leq e^{-\frac{t^2}{2\left(np + \frac{t}{3} \right)}} + e^{-\frac{t^2}{2np}}.
\end{displaymath}
Setting $t = 3 \sqrt{np \ln{n}}$ and noting that $t = o(np)$, we then have that $\Pr(||L_i| - np| \geq 3 \sqrt{np \ln{n}}) \leq e^{-4 \ln{n}}$ and the lemma follows from Boole's inequality.

For the second part, fix a vertex $v$. Notice that $|S_v|$ is a binomial random variable with parameters $m, p$, i.e. $|S_v| \sim {\cal B}(m, p)$. By Chernoff bounds, for any $\delta \geq 0$, we have that

\begin{displaymath}
\Pr(|S_v| > (1+\delta) mp) < \left( \frac{e^\delta}{(1+\delta)^{(1+\delta)}} \right)^{mp}.
\end{displaymath}
Setting $\delta = \frac{1}{mp} (3 \sqrt{mp \ln{m}} + \ln{n})$ and using Boole's inequality we get the desired result.

$\hfill \Box$

Notice that the above lemma provides a lower bound on the clique number. However, a clique in $G_{n, m, p}$ can be formed by combining more than one label. Clearly, a clique $Q$ which is not formed by a single label will need at least 3 labels, since 2 labels cannot cover all the edges needed for $Q$ to be a clique. In the discussion below, we will provide a much larger lower bound on the number of labels needed to form a clique $Q$ of size $|Q| \sim np$ which is not formed by a single label. The following definition will be useful.

\begin{definition}
Denote by $A_{y, x}$ the event that there are two disjoint sets of vertices $V_1, V_2 \subset V$, where $|V_1| = y$ and $|V_2| = x$ such that the following hold:

\begin{enumerate}

\item All vertices in $V_1$ have chosen some label $l_0$, i.e. $l_0 \in \cap_{u \in V_1} S_u$.

\item None of the vertices in $V_2$ has chosen $l_0$, i.e. $l_0 \notin \cup_{v \in V_2} S_v$.

\item Every vertex in $V_1$ is connected to every vertex in $V_2$.

\end{enumerate}  
\end{definition}

As a warm-up, we prove the following technical lemma, which is a first indication that in a $G_{n, m, p}$ graph, whp we cannot have $y$ too large and $x$ too small at the same time. This lemma will also be used as a starting step in the proof of our main theorem. 

\begin{lemma} \label{lemma1vertexout}
Let $G_{n, m, p}$ be a random instance of the random intersection graphs model with $m = n^{\alpha}, 0<\alpha<1$ and $p = \Omega\left(\sqrt{\frac{1}{nm}} \right)$ and $mp^2 = O(1)$. Then, for any $y \geq np \left(1 - o\left(\frac{1}{\ln{n}} \right) \right)$, $\Pr(A_{y, 1}) = o(1)$.
\end{lemma}
\proof Fix a particular label $l_0$, a subset $V_1$ of the vertices having chosen $l_0$ (i.e. $V_1 \subset L_{l_0}$) and a vertex $v \notin L_{l_0}$. The probability that $v$ is connected to all vertices in $V_1$ is exactly 

\begin{equation}
p(V_1, v) \stackrel{def}{=} \sum_{k=1}^{m-1} {m-1 \choose k} p^k (1-p)^{m-k-1} (1 - (1-p)^k)^y.
\end{equation}
Indeed, $p^k (1-p)^{m-k-1}$ is the probability that $v$ has chosen $k$ specific labels different from $l_0$ and $1 - (1-p)^k$ the probability that a specific vertex in $V_1$ has chosen at least one of those labels (so that it is connected to $v$). 

By Boole's and Markov's inequality we then have that 

\begin{equation}
\Pr(A_{y, 1}) \leq m {|L_{l_0}| \choose y} (n-|L_{l_0}|) p(V_1, v) 
\end{equation} 
By Lemma \ref{lemmaconcentration}, for any vertex $v$, we have that $|S_v| \leq (1+o(1))mp + \ln{n}$ whp. Since $(1 - (1-p)^k)^y$ is increasing in $k$ and also ${m-1 \choose k} p^k (1-p)^{m-k-1}$ is maximum around $mp$, we conclude that the maximum of ${m-1 \choose k} p^k (1-p)^{m-k-1} (1 - (1-p)^k)^y$ for $k \in \{1 \ldots (1+o(1))mp\}$ is attained at some index $k' = (1+o(1))mp$. Therefore,

\begin{eqnarray}
\Pr(A_{y, 1}) & \leq & m^2 n { |L_{l_0}| \choose y} {m-1 \choose k'} p^{k'} (1-p)^{m-k'-1} (1 - (1-p)^{k'})^y +o(1) \\
& \leq & m^2 n { |L_{l_0}| \choose y} (1 - (1-p)^{k'})^y +o(1)
\end{eqnarray}
where the $o(1)$ term corresponds to the error term from Lemma \ref{lemmaconcentration}. Using now the fact that (by the expansion of the natural logarithm) $(1-p)^\frac{1}{p} = e^{\frac{1}{p} \ln{(1-p)}} = e^{- \sum_{j=1}^\infty \frac{p^{j-1}}{j}} \geq e^{-1 - \sum_{j=2}^\infty p^{j-1}} = e^{-1 - \frac{p}{1-p}} \geq e^{-1.1}$, for any $p \to 0$, we have that

\begin{eqnarray}
\Pr(A_{y, 1}) & \leq & m^2 n { |L_{l_0}| \choose y} (1 - e^{-2mp^2})^y +o(1) \\
& = & m^2 n { |L_{l_0}| \choose |L_{l_0}| - y} (1 - e^{-2mp^2})^y +o(1) \\
& \leq & m^2 n (|L_{l_0}|)^{|L_{l_0}| - y} (1 - e^{-2mp^2})^y +o(1).
\end{eqnarray}
For any $y \geq |L_{l_0}| \left(1 - o\left(\frac{1}{\ln{n}} \right) \right)$, we then have that $\Pr(A_{y, 1}) \to 0$. But by Lemma \ref{lemmaconcentration} we have also that $|L_{l_0}| \leq np \left(1 + o\left(\frac{1}{\ln{n}} \right) \right)$, which completes the proof. 

$\hfill \Box$

The above lemma has the following alternative interpretation, which will be useful in the sequence:

\begin{corollary} \label{interpretationlemma1vertexout}
Let $G_{n, m, p}$ be a random instance of the random intersection graphs model with $m = n^{\alpha}, 0<\alpha<1$, $p = \Omega\left(\sqrt{\frac{1}{nm}} \right)$ and $mp^2 = O(1)$. Let also $Q$ be a clique in $G_{n, m, p}$ that is not formed by a single label and also $|Q| \sim np$. If $l_0 \in {\cal M}$ is any label chosen by some vertex $v \in Q$, then there is a positive constant $c' < \frac{1-\alpha}{2}$, such that whp there are at least $n^{c'}$ vertices in $Q$ that have not chosen $l_0$.
\end{corollary}
\proof Notice that, by assumption, $np = \Omega(n^{\frac{1-\alpha}{2}})$. Therefore, for any positive $c' < \frac{1-\alpha}{2}$, we have that $n^{c'} = o\left(\frac{np}{\ln^2{n}} \right)$. The result then follows by Lemma \ref{lemma1vertexout}.

$\hfill \Box$

We now strengthen the above analysis by using the following simple observation: For a set of vertices $V_2$ and $k \geq 2$, let $S_{V_2}^{(k)} \subseteq {\cal M}$ denote the set of labels that have been chosen by at least $k$ of the vertices in $V_2$. Then the probability that every vertex of a set of vertices $V_1$ is connected to every vertex in $V_2$ is at most

\begin{eqnarray}
p(V_1, V_2) & \leq & \left( |S_{V_2}^{(2)}| p + (1-p)^{|S_{V_2}^{(2)}|} \prod_{v \in V_2} \left(1 - (1-p)^{|S_v - S_{V_2}^{(2)}|} \right)\right)^y \\
& \leq & \left( |S_{V_2}^{(2)}| p + \prod_{v \in V_2} \left(1 - (1-p)^{|S_v|} \right)\right)^y \label{pv1v2}
\end{eqnarray}
Indeed, the first of the above inequalities corresponds to the probability that each vertex in $V_2$ either choses one of the labels shared by at least two vertices in $V_2$, or it is connected to all vertices in $V_2$ by using labels chosen by exactly one vertex in $V_2$.

\begin{lemma} \label{lemmamanyverticesout}
Let $G_{n, m, p}$ be a random instance of the random intersection graphs model with $m = n^{\alpha}, 0<\alpha<1$, $p = \Omega\left(\sqrt{\frac{1}{nm}} \right)$ and $mp^2 = O(1)$. Let also $x = \frac{1}{p^\epsilon}$, for some positive constant $\epsilon <1$ that can be as small as possible. Then, for any $y \geq  np^{1+c}$, where $0< c < \frac{1-\alpha}{1+\alpha}$ is a constant, we have $\Pr(A_{y, x}) = o(1)$.
\end{lemma}
\proof Fix a set $V_2$ of $x$ vertices. We first give an upper bound on the size of $S_{V_2}^{(2)}$. Towards this end, let $X = |S_{V_2}^{(2)}|$ and notice that $X$ is binomially distributed with parameters $m, \hat{p} = 1 - (1-p)^x - xp (1-p)^{x-1}$. Since, by assumption $xp \to 0$, we have that $\hat{p} \leq \frac{x^2 p^2}{2}$. Therefore $X$ is stochastically dominated by a binomial random variable $Y \sim {\cal B}\left( m, \frac{x^2 p^2}{2}\right)$. 

By Chernoff bounds we then have, for any $t \geq 0$, 

\begin{equation}
\Pr\left(X > \frac{mx^2 p^2}{2} +t \right) \leq e^{-\frac{t^2}{2\left(\frac{mx^2 p^2}{2} + \frac{t}{3} \right)}}
\end{equation}
Set $t = \frac{1}{p^{2\epsilon+\epsilon'}}$, where $\epsilon'$ is a positive constant that can be as small as possible. Since $mp^2 = O(1)$, we have that $t = \omega\left( \frac{mx^2 p^2}{2} \right)$. By Boole's inequality then, the probability that there is a subset $V_2$ of $x$ vertices that has $|S_{V_2}^{(2)}| > \frac{mx^2 p^2}{2} + \frac{x^2}{p^{\epsilon'}}$ is at most

\begin{equation} \label{errorsv22}
n^x e^{-\frac{1}{3 p^{2\epsilon+\epsilon'}}} = o(1). 
\end{equation}

Now that we have an upper bound on the size of $S_{V_2}^{(2)}$ that holds whp, notice that by the second part of Lemma \ref{lemmaconcentration} and the fact that $mp^2 = O(1)$, whp we have

\begin{equation}
\prod_{v \in V_2} \left(1 - (1-p)^{|S_v|} \right) \leq \frac{1}{2^{\Theta(x)}} = o(|S_{V_2}^{(2)}| p).
\end{equation}
Therefore, by (\ref{pv1v2}), we have that $p(V_1, V_2) \leq (2|S_{V_2}^{(2)}| p)^{|V_1|}$. By Boole's and Markov's inequality we then have that 

\begin{eqnarray}
\Pr(A_{y, x}) & \leq & m {|L_{l_0}| \choose y} n^{x} p(V_1, V_2) \\
& \leq & m {|L_{l_0}| \choose y} n^{x} (2|S_{V_2}^{(2)}| p)^{y} + o(1) \\
& \leq & m {|L_{l_0}| \choose y} n^{x} \left( 2p^{1-2\epsilon-\epsilon'} \right)^y + o(1)
\end{eqnarray} 
where the $o(1)$ term corresponds to the error terms from Lemma \ref{lemmaconcentration} and equation (\ref{errorsv22}). Using now the first part of Lemma \ref{lemmaconcentration} and an upper bound for the binomial coefficient we have

\begin{eqnarray}
\Pr(A_{y, x}) & \leq & m \left( \frac{8 np}{y} \right)^y n^{x} \left(p^{1-2\epsilon-\epsilon'} \right)^y + o(1).
\end{eqnarray} 
Setting $y = np^{1+c}$, for any positive constant $c < \frac{1-\alpha}{1+\alpha}$, we have ($y \geq 1$ and also) that $\Pr(A_{y, x}) = o(1)$. This completes the proof.

$\hfill \Box$

Lemma \ref{lemmamanyverticesout} has the following interpretation:

\begin{corollary} \label{interpretationlemmamanyverticesout}
Let $G_{n, m, p}$ be a random instance of the random intersection graphs model with $m = n^{\alpha}, 0<\alpha<1$, $p = \Omega\left(\sqrt{\frac{1}{nm}} \right)$ and $mp^2 = O(1)$. Let also $Q$ be a clique in $G_{n, m, p}$ that is not formed by a single label and also $|Q| \sim np$. Then whp, for any label $l_0 \in {\cal M}$, we have that $|Q \cap L_{l_0}| \leq np^{1+c}$, where $0< c < \frac{1-\alpha}{1+\alpha}$ is a constant. 

In particular, if $Q$ is not formed by a single label, then whp it is formed by at least $\frac{1}{p^c}$ distinct labels.
\end{corollary}
\proof By Corollary \ref{interpretationlemma1vertexout}, if $Q$ is not formed by a single label, then given any label $l_0 \in {\cal M}$ which is chosen by some vertex $v \in Q$, there is a positive constant $c' < \frac{1-\alpha}{2}$, such that whp there are at least $n^{c'}$ vertices in $Q$ that have not chosen $l_0$. Therefore, we can apply Lemma \ref{lemmamanyverticesout} using any $\epsilon < \frac{2c'}{1+\alpha}$. More specifically, for any such $\epsilon$ we have $\Pr(A_{np^{1+c}, \frac{1}{p^{\epsilon}}}) = o(1)$. 

Additionally, this implies that whp if $Q$ is not formed by a single label, it needs at least $\frac{np}{np^{1+c}} = \frac{1}{p^c}$ distinct labels. This is also a lower bound on the number of labels needed by a vertex $v$ in order to connect to all vertices in $Q$.

$\hfill \Box$

Before presenting the proof of our main theorem, we prove the following useful lemma, which states that if a large clique is not formed by a single label, then it must contain a quite large clique $Q'$ whose edges are formed by distinct labels.

\begin{lemma} \label{lemmainnerindependentclique}
Let $G_{n, m, p}$ be a random instance of the random intersection graphs model with $m = n^{\alpha}, 0<\alpha<1$, $p = \Omega\left(\sqrt{\frac{1}{nm}} \right)$ and $mp^2 = O(1)$. Let also $Q$ be any clique in $G_{n, m, p}$ that is not formed by a single label and also $|Q| \sim np$. Then whp, $Q$ contains a clique $Q'$ whose edges are formed by distinct labels and whose size is at least $p^{-\frac{c}{2}}$, for any positive constant $c < \frac{1-\alpha}{1+\alpha}$.
\end{lemma}
\proof Let $Q'$ be a subset of $Q$ which is maximal with respect to the following property ${\cal P}$: ``to each pair of vertices $u \neq v$ in $Q'$ we can assign a distinct label $l$, such that $l \in S_u \cap S_v$''. 

Consider now the set of vertices $W = \{w: S_w \cap S_{Q'}^{(2)} \neq 0\}$, namely the set of vertices that share a label with at least 2 vertices in $Q'$ (note that $Q' \subseteq W$, because every pair of vertices in $Q$ is connected). Since $Q'$ is maximal, the set $Q - W$ must be the empty set. Indeed, if $z \in Q - W$, then (baring in mind that $Q$ is a clique) $z$ can be connected to each vertex in $Q'$ using distinct labels, which are also different from those already used to connect pairs of vertices in $Q'$. Therefore, $Q' \cup \{z\}$ would also have property ${\cal P}$, which contradicts the maximality of $Q'$.   

By Corollary \ref{interpretationlemmamanyverticesout} now, we have that $|W| \leq |S_{Q'}^{(2)}| np^{1+c}$, where $0< c < \frac{1-\alpha}{1+\alpha}$ is a constant. Furthermore, by equation (\ref{errorsv22}), we have that $|S_{Q'}^{(2)}| \leq \frac{m|Q'|^2 p^2}{2} + \frac{|Q'|^2}{p^{\epsilon'}}$ whp, for any $\epsilon'>0$ that can be as small as possible. Combining the above, and since $mp^2 = O(1)$, we have that 

\begin{equation}
|W| \leq \frac{np^{1+c} |Q'|^2}{(1+o(1))p^{\epsilon'}}.
\end{equation}
Consequently, the requirement $Q - W = \emptyset$ translates to 

\begin{equation}
|Q| - \frac{np^{1+c} |Q'|^2}{(1+o(1))p^{\epsilon'}} \leq 0
\end{equation} 
or equivalently

\begin{equation}
|Q'| \geq \sqrt{\frac{|Q|}{(1+o(1)) np^{1+c - \epsilon'}}}.
\end{equation}
Baring in mind that $|Q| \sim np$ and that $\epsilon'>0$ can be as small as possible, this completes the proof.

$\hfill \Box$

We now present our main theorem.

\begin{theorem}[Single Label Clique Theorem] \label{singlelabelclique}
Let $G_{n, m, p}$ be a random instance of the random intersection graphs model with $m = n^{\alpha}, 0<\alpha<1$ and $mp^2 = O(1)$. Then whp, any clique $Q$ of size $|Q| \sim np$ in $G_{n, m, p}$ is formed by a single label. In particular, the maximum clique is formed by a single label.
\end{theorem}
\proof We first note that, as discussed in the beginning of section \ref{sectionclique}, when $p = o\left(\sqrt{\frac{1}{nm}} \right)$, by slightly modifying an argument of \cite{BTU08} (in particular Lemma 5 there), we can see that $G_{n, m, p}$ almost surely has no cycle of size $k \geq 3$ whose edges are formed by $k$ distinct labels. Therefore, the maximum clique of $G_{n, m, p}$ when $p = o\left(\sqrt{\frac{1}{nm}} \right)$, is formed by exactly one label and our theorem holds. Consequently, we will assume w.l.o.g. for the remainder of the proof that $p = \Omega\left(\sqrt{\frac{1}{nm}} \right)$. 

Let $Q$ be a clique of size $|Q| \sim np$ in $G_{n, m, p}$. By Lemma \ref{lemmainnerindependentclique}, if $Q$ is not formed by a single label, then $G_{n, m, p}$ must contain a clique $Q'$ whose edges are formed by distinct labels and whose size is at least $\beta \stackrel{def}{=} p^{-\frac{c}{2}}$, for any positive constant $c < \frac{1-\alpha}{1+\alpha}$. By Markov's inequality, the probability that such a $Q'$ exists in $G_{n, m, p}$ is at most

\begin{equation}
n^{\beta} \prod_{k=1}^{\beta-1} {m \choose \beta-k} p^{2(\beta-k)}.
\end{equation} 
Indeed, we can choose the vertices in $Q'$ arranged in a line in at most $n^{\beta}$ ways. Then we can select the labels needed for the $k$-th vertex to connect to all vertices to its right in at most ${m \choose \beta-k}$ ways and each such label must be chosen by the $k$-th vertex, as well as another vertex to its right (hence the term $p^{2(\beta-k)}$ in the product). Upper bounding the binomial coefficients in the above and using the fact $mp^2 = O(1)$, we get

\begin{eqnarray}
\Pr\{\textrm{$Q'$ exists in $G_{n, m, p}$}\} & \leq & n^{\beta} \prod_{k=1}^{\beta-1} \left( \frac{e m}{\beta-k}\right)^{\beta-k} p^{2(\beta-k)} \\
& \leq & n^{\beta} \prod_{k=1}^{\beta-1} \left( \frac{\Theta(1)}{\beta-k}\right)^{\beta-k} = e^{\beta \ln{n} + O(\beta^2)} \prod_{k=1}^{\beta-1} \left( \frac{1}{\beta-k}\right)^{\beta-k} \\
& \leq & e^{\beta \ln{n} + O(\beta^2)} \prod_{k= \frac{\beta}{2}}^{\beta-1} \left( \frac{1}{\beta-k}\right)^{\beta-k} \\ 
& \leq & e^{\beta \ln{n} + O(\beta^2)} \prod_{k= \frac{\beta}{2}}^{\beta-1} \left( \frac{2}{\beta}\right)^{\beta-k} = e^{\beta \ln{n} + O(\beta^2)} \frac{1}{\beta^{\Theta(\beta^2)}} = o(1).
\end{eqnarray}
Therefore, whp $Q'$ does not exist in $G_{n, m, p}$, which completes the proof.

$\hfill \Box$

Notice that, by Theorem \ref{singlelabelclique}, the maximum clique in $G_{n, m, p}$ with $m = n^{\alpha}, 0<\alpha<1$ and $mp^2 = O(1)$ must be one of the sets $L_l, l \in {\cal M}$. Therefore, the clique number of $G_{n, m, p}$ can be bounded using the first part of Lemma \ref{lemmaconcentration}. In particular

\begin{corollary} \label{maxcliquesize}
Let $G_{n, m, p}$ be a random instance of the random intersection graphs model with $m = n^{\alpha}, 0<\alpha<1$, $p = \Omega\left(\sqrt{\frac{1}{nm}} \right)$ and $mp^2 = O(1)$. Then, whp the maximum clique $Q$ of $G_{n, m, p}$ satisfies $|Q| \sim np$.
\end{corollary}


\section{Label Reconstruction} \label{labelreconstruction}

One of the implications of our main Theorem \ref{singlelabelclique} is that whp we can find the maximum clique in $G_{n, m, p}$ with $m = n^{\alpha}, 0<\alpha<1$ and $mp^2 = O(1)$ in polynomial time, just by looking at the associated bipartite graph $B_{n, m, p}$. In the following algorithm, we denote by $L_i$ the set of neighbors of label $i \in {\cal M}$ in $B_{n, m, p}$, which can be determined in $O(n)$ time.

\vspace{0.5cm}

\begin{tabular}{|p{12cm}|}
\hline
\noindent \texttt{\textbf{Algorithm}} MAX-CLIQUE\_FROM\_LABELS \\
\noindent \texttt{\textbf{Input:}} $B_{n, m, p}$

\begin{enumerate}

\item Set $Q = \emptyset$;

\item \texttt{\textbf{for}} $i=1$ \texttt{\textbf{to}} $m$ \texttt{\textbf{do}}

\% Check if the clique induced by label $i$ is larger \%

\item \quad \texttt{\textbf{if}} $|L_i| > |Q|$ \texttt{\textbf{then}} set $Q = L_i$; \texttt{\textbf{endfor}}

\item \texttt{\textbf{Output}} $Q$;
\end{enumerate} \\
\hline 
\end{tabular}

\vspace{0.5cm}

By Theorem \ref{singlelabelclique}, when $m = n^{\alpha}, 0<\alpha<1$ and $mp^2 = O(1)$, Algorithm MAX-CLIQUE\_FROM\_LABELS returns the maximum clique of $G_{n, m, p}$ whp, in $O(nm)$ time. Therefore, the randomness of the model works in our favor for this case. Indeed, since any graph can be written as an intersection graph with at most ${n \choose 2}$ labels, the problem of finding a maximum clique in a graph, given its label representation remains NP-complete. Furthermore, it remains hard even when the intersection number is $n^{\alpha}, 0< \alpha<1$ unless the exponential time hypothesis fails (see e.g. \cite{CHKX06}).

This leads to the following natural question: Could one infer any information about the structure of the associated bipartite graph when provided with $G_{n, m, p}$ (i.e. the vertices and the edges of the graph)? Notice here that a graph $G_{n, m, p}$ can correspond to more than one associated bipartite graphs. However, we show here that the problem of finding the associated bipartite graph given $G_{n, m, p}$ and the actual values of $m, n$ and $p$ is \emph{solvable} whp when the number of labels is less than the number of vertices; namely, the maximum likelihood estimation method will provide a unique solution (up to permutations of the labels). More specifically, if ${\cal B}_{n, m, p}$ is the set of non-isomorphic associated bipartite graphs that give rise to $G_{n, m, p}$, then there is some $B^* \in {\cal B}_{n, m, p}$ such that $\Pr(B^*|G_{n, m, p}) > \Pr(B|G_{n, m, p})$, for all ${\cal B}_{n, m, p} \ni B \neq B^*$. 

\begin{theorem} \label{uniquelabelrepresentation}
Let $G_{n, m, p}$ be a random instance of the random intersection graphs model with $m = n^{\alpha}, 0<\alpha<1$, $p = \Omega\left(\sqrt{\frac{1}{nm}} \right)$ and $mp^2 = O(1)$. Then, whp the bipartite graph $B_{n, m, p}$ associated to $G_{n, m, p}$ is uniquely determined, up to permutations of the labels.
\end{theorem}
\proof Let $L_{i}$ denote the set of vertices that have chosen label $i \in {\cal M}$. Given $G_{n, m, p}$, we will refer to $\{L_{i}, i \in {\cal M}\}$ as its \emph{label representation}. Notice then that the associated bipartite graph is uniquely defined by the label representation. 

Suppose now for the sake of contradiction that $\{L_{i}^{(1)}, i \in {\cal M}\}$ and $\{L_{i}^{(2)}, i \in {\cal M}\}$ are two distinct label representations (up to permutations of the labels) of $G_{n, m, p}$, where $m = n^{\alpha}, 0<\alpha<1$, $p = \Omega\left(\sqrt{\frac{1}{nm}} \right)$ and $mp^2 = O(1)$. Notice that, by the first part of Lemma \ref{lemmaconcentration}, whp both of these label representations should satisfy $|L_{i}^{(\xi)}| \sim np$, for any $i \in {\cal M}$ and $\xi = 1, 2$. 

Notice then that there must be a label $l$, such that $L_{l}^{(1)} \notin \{L_{i}^{(2)}, i \in {\cal M}\}$, i.e. the clique induced by label $l$ can be edge covered by more than one other cliques of size asymptotically equal to $np$. However, by Theorem \ref{singlelabelclique}, whp no clique $Q$ of size $|Q| \sim np$ can be formed by more than one labels, which contradicts the assumption that $L_{l}^{(1)} \notin \{L_{i}^{(2)}, i \in {\cal M}\}$. 

Consequently, $\{L_{i}^{(1)}, i \in {\cal M}\}$ and $\{L_{i}^{(2)}, i \in {\cal M}\}$ must be similar, up to permutations of the labels, i.e. $L_l^{(1)} \in \{L_{i}^{(2)}, i \in {\cal M}\}$, for every $l \in {\cal M}$. This completes the proof.

$\hfill \Box$ 

Notice that the uniqueness of the bipartite graph can also be proved in the case where $p = o\left(\sqrt{\frac{1}{nm}} \right)$. Indeed, in this case $G_{n, m, p}$ almost surely has no cycle of size $k \geq 3$ whose edges are formed by $k$ distinct labels (see also the beginning of Section \ref{sectionclique}). Therefore, every clique of size at least 3 is formed by a single label and so the proof of Theorem \ref{uniquelabelrepresentation} applies in this (sparser) case also.

\section{Conclusions} \label{conclusions}

In this work, we studied the maximum clique problem by relating it to the intersection number of the input graph. In particular, we first proved that if the intersection number of the graph $G$ is sufficiently small, then a simple algorithm can find a maximum clique in $G$ in polynomial time. We then considered random instances of the random intersection graphs model as input graphs. In particular, by proving the Singe Label Clique Theorem, we provided new, more general and asymptotically tight bounds for the clique number of $G_{n, m, p}$ when $m = n^{\alpha}, \alpha<1$. We also claim that our proof carries over for $\alpha<2$, provided there is a lower bound on $p$ (in particular, we claim that our analysis can be applied also for $mp^2 = \Theta(1)$). One of the consequences of our theorem is that we can use the label representation of $G_{n, m, p}$ to find a maximum clique in polynomial time whp. This raised the question of whether we could reconstruct the label choices of the vertices in $G_{n, m, p}$ given only the graph structure. We proved here that the label reconstruction problem is solvable whp when the number of labels is less than the number of vertices. Finding efficient algorithms for constructing such a label representation is left as an open problem for future research. In view of the equivalence results between random intersection graphs and Erd\H{o}s-R\'enyi random graphs, we expect that our work will shed light also in the problem of finding maximum cliques for input graphs generated by the latter model.

\end{document}